\documentclass[%
preprint,
superscriptaddress,
 amsmath,amssymb,
 aps,
prb,
]{revtex4-2}

\usepackage[utf8]{inputenc}
\usepackage{graphicx}
\usepackage{subfigure}
\usepackage[inkscapelatex=false]{svg}          
\usepackage[justification=raggedright,singlelinecheck=false]{caption}
\usepackage{subcaption}   

\usepackage{bm}

\begin{document}

\title{Vibrational frequencies and stark tuning rate with continuum electro-chemical models and grand canonical density functional theory}

\author{Mouyi Weng}
\affiliation{ Theory and Simulation of Materials (THEOS), École Polytechnique Fédérale de Lausanne (EPFL), CH-1015 Lausanne, Switzerland}
\author{Nicéphore Bonnet}
\affiliation{ Theory and Simulation of Materials (THEOS), École Polytechnique Fédérale de Lausanne (EPFL), CH-1015 Lausanne, Switzerland}
\author{Oliviero Andreussi}
\affiliation{ Department of Chemistry and Biochemistry, Boise State University, Boise, Idaho 83725, USA}
\author{Nicola Marzari}
 \email{nicola.marzari@epfl.ch}
 \affiliation{ Theory and Simulation of Materials (THEOS), École Polytechnique Fédérale de Lausanne (EPFL), CH-1015 Lausanne, Switzerland}

\date{\today}

\begin{abstract}

Simulating electrochemical interfaces using density functional theory (DFT) requires incorporating the effects of electrochemical potential. The electrochemical potential acts as a new degree of freedom that can effectively tune DFT results as electrochemistry does. Typically, this is implemented by adjusting the number of electrons on the solid surface within the Kohn-Sham (KS) equation, under the framework of an implicit solvent model and the Poisson-Boltzmann equation (PB equation), thereby modulating the potential difference between the solid and liquid. These simulations are often referred to as grand-canonical or fixed-potential DFT calculations. To apply this additional degree of freedom, Legendre transforms are employed in the calculation of free energy, establishing the relationship between the grand potential and the free energy. Other key physical properties, such as atomic forces, vibrational frequencies, and Stark tuning rates, can be derived based on this relationship rather than directly using Legendre transforms. This paper begins by discussing the numerical methodologies for the continuum model of electrolyte double layers and grand-potential algorithms. We then show that atomic forces under grand-canonical ensemble match the Hellmann-Feynman forces observed in canonical ensemble, as previously established. However, vibrational frequencies and Stark tuning rates exhibit distinct behaviors between these conditions. Through finite displacement methods, we confirm that vibrational frequencies and Stark tuning rates exhibit differences between grand-canonical and canonical ensembles. 

\end{abstract}

\maketitle

\section{\label{sec:level1}Introduction}
Simulating electrochemical interfaces using density functional theory (DFT) requires incorporating the effects of electrochemical potential\cite{Resasco2022_perspective, Melander2020, KopacLautar2020_Mgion_battery, Bhandari2021}. The computational hydrogen electrode (CHE) approach is a widely used method to incorporate electrochemical potential effects into reaction energy calculations\cite{Valdes2008_Nors_CHE1, Nørskov2004_CHE2}. This approach is based on the relationship $\Delta G=\Delta N_e \times U$, which follows from Janak’s theorem \cite{Janak1978} stating that the total energy change is given by the electrochemical potential times the change in the number of electrons. Notably, this approach relies on the approximation that $\Delta N_e$ is sufficiently small, the reaction surface is infinitely large, or the density of reaction sites is sufficiently low, all three of which are equivalent conditions ensuring that the electronic structure remains unchanged. While the CHE method performs well for many chemical reactions, it introduces errors when the electronic structure changes significantly before and after the variation in the number of electrons\cite{Gao2020_LWW_GC}. An alternative is to use a grand-canonical DFT approach, which differs from the CHE approach. In grand-canonical DFT, the electrochemical potential between the solid and liquid phases is fixed, and a hypothetical reservoir of electrons adjusts the potential difference between the liquid and solid bulk phases. This approach directly accounts for the changes in the electron number and their effect on the electronic structure of the entire system. This method is also referred to as fixed-potential, fixed-Fermi-energy, or grand-potential DFT calculations. By contrast, conventional DFT, where the number of electrons remains constant, is referred to as fixed-electron-number or canonical DFT calculations.

With the grand-potential DFT simulation approach, many studies have investigated various electrochemical phenomena, including the  electrochemical double layer\cite{Chen2018_doublelayer}, CO2 reduction\cite{DOI:10.1038/s41467-018-05544-3}, defect and adsorption energy\cite{Beinlich2022}, band-alignment calculation\cite{Hormann2019_surface_water}, Hydrogen Peroxide Reduction Reaction\cite{Rojas2020}, Oxide electro-Reduction process on metal surface\cite{Pfisterer2020}, electrosorption on metal surface\cite{Hormann2020}, (Oxygen Evolution and Reduction) OER\cite{Karmodak2022} and Hydrogen Evolution Reaction (HER) catalysis process\cite{doi:10.1021/acsenergylett.9b02689}on two dimensional materials and single atom catalysis\cite{Hossain2020_Single_atom}. To apply electrode potential in DFT calculations, two main approaches are commonly employed. The first method fixes the number of electrons on the solid surface. By performing multiple canonical DFT calculations, surface electrochemical potential can be converged to match experimental results\cite{doi:10.1063/1.5054580}. The second method directly imposes the desired potential between the solid and liquid phases within a modified DFT self-consistent loop. A specially designed charge mixing method in Kohn-Sham (KS) equation solver then converges to a charged state on the solid surface that satisfies the input potential\cite{doi:10.1063/1.4978411}. This approach has been recently implemented in many open-source codes like JDFTx\cite{doi:10.1063/1.4978411}, CP2K\cite{Chai2024}, GPAW\cite{Melander2024} and the effective screening medium (ESM) method in Quantum ESPRESSO\cite{Haruyama2018}. For a fully converged self-consistent calculation on a given atomic structure under the same potential, grand-canonical and canonical calculations yield the same electronic structure. The relationship between the grand potential in the grand-canonical ensemble and the Helmholtz free energy in the canonical ensemble can be derived using a Legendre transform. However, for other properties, such as vibrational frequencies, and Stark tuning rates, these two methods do not necessarily yield the same results. 

In this paper, we first describe the continuum model for the solid-liquid interface, which incorporates an implicit solvent and the Poisson-Boltzmann equation. We then introduce the fixed-potential approach for achieving self-consistency in the self-consistent solver. As the core contribution of this work, we demonstrate that atomic forces remain consistent between the canonical and grand-canonical ensembles, in agreement with previous studies\cite{Duan2021_GC_forces}. We further derive that vibrational frequencies and Stark tuning rates exhibit intrinsic differences between the two ensembles. To verify this theoretical prediction, we employ finite difference calculations. Based on our analytical derivations and finite-difference calculations, we show that there is a difference between grand-canonical vibrational frequencies and canonical vibrational frequencies, moreover, this difference can be accurately calculated using our derived expressions. The consistency between the directly computed grand-canonical results and those obtained via transformation from canonical data further validates our theoretical framework.

\section{Continuum Models and Grand-Canonical Self-Consistent Methods}

\subsection{Continuum Models of Solvents and Electrolytes for Electro-Chemical Simulations}

The simulation of the electrochemical environment at the atomic level is one of the most sophisticated problems in DFT calculations\cite{Groß2022,Ringe2022}. Factors such as dielectric screening from the solvent\cite{Tomasi2005_chemrev_solvent, Marenich2007_SM8_solvationmodel, doi:10.1063/1.3676407}, electrolyte double layer\cite{Jinnouchi2008_PBe_solver, Nattino2019, doi:10.1063/1.4939125}, electrode potential\cite{Duan2021_GC_forces, Vijay2022_NEB_GC, doi:10.1063/1.5054580, Gao2020_LWW_GC} and even the chemical reactivity or hydrogen-bond formation on interfaces\cite{Xi2022_LWW_hybrid_solvation, Groß2022_chemreview, Hormann2019_surface_water} should be considered in the characterization of wet electrified interfaces. Although most DFT studies relying on plane-wave basis sets and periodic boundary conditions focus on neutral interfaces, handling charged surfaces is also of paramount importance for the simulation of electrochemical environments \cite{Chan2015_surface_charge_neb, Dabo2008_parabolic, Otani2006_esm}. 

One approach to include all these factors in DFT simulations is to setup an extensive detailed system that includes every component with explicit atomistic details, followed by molecular dynamics (MD) simulations to statistically average the contribution of the disordered liquid/electrolyte components on the properties of interest \cite{Le2017_chengjun_PRL, Groß2022_explicit_water_md, Ong2011_Oliviero_MD_solvent, Sundararaman2022}. In this approach, significant computational cost is spent on simulating atoms far from the interface and achieving time-averaged statistical convergence. Although significant progress has been achieved in the last decade in the development of machine learning schemes to accelerate MD simulations \cite{Zeng2023_ml_surface_water}, the variety and complexity of the components involved in wet electrified interfaces still limit the accuracy of explicit simulations when compared to experimental results\cite{Wang2021_zss_nature}. 

An alternative approach to capture the complexity of solid-liquid interfaces in DFT simulations is to convert some challenging components, particularly those requiring statistical sampling, into continuum embeddings. This hierarchical approach treats the effects from explicit molecular systems (e.g., liquids and electrolytes) using a statistically averaged implicit continuum medium. Thus, this approach focuses most computational efforts on the system responsible for the properties of interest, while avoiding spending computational resources on atoms and molecules far from the interface. When applied to all statistically disordered components in the simulations, the continuum approach can also reduce or remove the need for MD sampling, thus significantly lowering the computational costs and challenges associated with MD simulations. Most definitions of continuum solvation models rely on a more or less extended set of parameters, as required, in particular, for defining the boundaries of the embedding media and tuning the strengths of the interactions between the DFT and the continuum systems. Recent models have shown that a minimal set of empirically tuned parameters can provide results near or below chemical accuracy on several properties of interest, including solvation energy for neutral\cite{doi:10.1063/1.3676407, Hille2019_SCCS_Parameters} and charged systems\cite{doi:10.1063/1.4832475}, surface capacity\cite{doi:10.1063/1.4939125, Nattino2019}, surface energy\cite{doi:10.1063/1.5054580}, and electro-adsorption energy as a function of pH and electrode potential\cite{Hormann2020}. In this paper, we rely on the self-consistent continuum solvation (SCCS) model of Andreussi et al.\cite{doi:10.1063/1.3676407}, which has been extensively parametrized and tested in recent years. For the sake of introducing the notation and identifying the key parameters of continuum models, in the following we summarize the key equations of SCCS.  

The calculation of electrostatic interactions in standard DFT simulations requires solving the Poisson equation in vacuum
\begin{equation}
  \nabla^2\Phi^{tot}(\bold r)=-4\pi \rho^{solute}(\bold r)\label{poisson}  
\end{equation}
to identify the electrostatic potential, $\Phi^{tot}(\mathbf{r})$, generated by the solute charge density, $\rho^{solute}(\mathbf{r})$. In the presence of a solvent, a non-homogeneous dielectric medium, $\epsilon(\mathbf{r})$, is introduced in the simulation cell, and the calculation of the electrostatic potential requires solving the more challenging generalized Poisson equation
\begin{equation}
    \nabla\cdot\epsilon(\bold r)\nabla\Phi^{tot}(\bold r)=-4\pi \rho^{solute}(\bold r). \label{poisson_solvation}  
\end{equation}
The dielectric function is designed to switch from a value of 1 (vacuum), where the DFT system's degrees of freedom are present, to the bulk dielectric permittivity of the solvent, $\epsilon_0$, in the bulk of the continuum medium and far from the system's interface. The SCCS model introduces a smooth transition between these two asymptotic values and models it based on the electronic charge density of the solute, so that the interface itself can readjust following the electrons' optimization in the self-consistent field (SCF) loop of DFT calculations. Namely, in SCCS 
\begin{equation}\label{frho}
    \epsilon(\rho(\bold r))=\begin{cases}
1&\ \  \rho^{elec}(\bold r)>\rho_{max,\epsilon}\\
f(\rho^{elec}(\bold r)) &\ \  \rho_{min,\epsilon}<\rho^{elec}(\bold r)<\rho_{max,\epsilon}
\\
\epsilon_0&\ \  \rho^{elec}(\bold r)<\rho_{min,\epsilon}
\end{cases}
\end{equation}
where the set of switching parameters, $\rho_{max,\epsilon}$ and $\rho_{min,\epsilon}$, controls the position of the interface with respect to the DFT system and it is usually tuned to reproduce specific solvation properties of the system. The connection between the dielectric function and the electronic density, $f(\rho)$, is expressed in SCCS as a smooth function that depends on the natural logarithm of the electronic density \cite{doi:10.1063/1.3676407}. This choice accounts for the exponential decay of the electronic density far from the atoms, thus ensuring that the polarization charge density is smooth enough to be well described on the same structured grids used for plane-wave DFT calculations.  

In addition to the dielectric screening in neural liquids, continuum models of electrolytes have been introduced within the SCCS framework \cite{Nattino2019}. Accounting for electrolyte ions requires extending the generalized Poisson equation into a nonlinear Poisson Boltzmann equation, 
\begin{equation}
    \nabla\cdot \epsilon(\bold r)\nabla\phi(\bold r)=-4\pi[\rho^{solute}(\bold r)+\rho^{ion}(\bold r)]\label{PBe}
\end{equation}
with
\begin{equation}
\rho^{ion}(\bold{r})=\gamma(\bold{r})\sum_i^m Z_ic_i[\phi](\bold{r}),
\label{gamma}
\end{equation}
in which the distribution of charge from each species $i$ of dissolved ions can be expressed in terms of their charges, $Z_i$, and their concentrations in the simulation cell, $c_i(\mathbf{r})$. A continuum interface function specific for the electrolyte ions, $\gamma(\mathbf{r})$, is also introduced, so as to ensure that the ions are only distributed in the region outside of the DFT substrate. This interface function has a definition similar to the dielectric function introduced in Eq~\eqref{frho}, but it relies on its own set of switching parameters, $\rho_{max,\gamma}$ and $\rho_{min,\gamma}$. 

Following the Gouy-Chapman description of the electrochemical diffuse layer, the concentrations of electrolyte ions can be expressed in terms of their bulk values, $c_i^\infty$, using a Boltzmann factor that involves the electrostatic potential energy 
\begin{equation}
   c_i[\phi](\bold r)=c_i^\infty\cdot exp\left(-\frac{Z_i e\phi(\bold r)}{k T} \right)
   \label{PBEci}
\end{equation}
Modifications to the above expression have been proposed in order to account for the finite size of electrolyte ions \cite{Borukhov2000}. The resulting modified Poisson-Boltzmann equation is known to produce smoother and more well-defined concentration profiles, avoiding diverging concentrations at high bias potentials.

The continuum description of the electrolyte charge distribution discussed above unlocks the use of periodic DFT simulations for charged slabs of materials, allowing the finite charge in the DFT system to be fully compensated by the charge in the continuum electrolyte and imposing the correct asymptotic behavior. When applied to noble metals in electrochemical environments, simulation results based on these continuum approaches show a good qualitative agreement with experimentally measured differential capacitance\cite{Nattino2019}.

However, the highly nonlinear nature of the Poisson Boltzmann equations can significantly hinder the numerical convergence of the electrostatic problem. For low electrostatic potentials or diluted electrolyte solutions, a linearized Poisson Boltzmann equation can be obtained, i.e., by only keeping the first term in the Taylor's expansion of $c_i[\phi](\bold r)$: 
\begin{equation}
   c_i[\phi](\bold r)=c_i^\infty\cdot\left(1-\frac{Z_i e\phi(\bold r)}{k T} \right)
\end{equation}
The linearized problem can be readily solved by mathematical methods such as the preconditioned conjugate-gradient method\cite{doi:10.1063/1.4939125}. In the following work, we used the linearized Poisson-Boltzmann equation to ensure numerical stability and convergence efficiency for all our simulations.

\subsection{Grand Potential Simulations}
By leveraging the continuum models described above, it is possible to perform a range of simulations in which the total charge on a materials substrate is varied systematically. The interface free energy of the resulting fixed-charge simulations can be recast into a grand canonical free energy that depends on the conjugated degree of freedom using a Legendre transform \cite{doi:10.1063/1.5054580}. If the degree of freedom that is varied in the fixed-charge simulations is the number of electrons, the Legendre-transformed quantity is a grand canonical free energy that depends on the applied bias potential, sometimes referred to as the grand potential. An alternative way to perform grand potential calculations is to set the electro-chemical potential before the calculation and let the SCF loop within DFT will automatically converge to the targeted potential while adjusting the number of electrons in the simulation. This method is often dubbed fixed-potential or fixed-fermi energy method. 

The fixed-charge approach does not require specific implementation strategies, apart from defining an interpolation on the fixed-charge results before performing the Legendre transform. The fixed-potential approach, instead, requires to adjust the SCF algorithm in DFT that allows to optimize the number of electrons constrained to produce the specified bias. We follow the approach of Sundararaman et al.\cite{doi:10.1063/1.4978411} for Pulay mixing and that of Haruyama et al.\cite{Haruyama2018} for implementing Broyden mixing in Quantum ESPRESSO\cite{Giannozzi_2017, Hagiwara2021_lgcscf}, and describe how the fixed-potential approach is realized.

Similar to other mixing methods used for non-linear optimization of the electronic density, Broyden mixing relies on the residue obtained at each SCF iteration $i$ in terms of the electronic density in input, $\rho^{elec}_{i,in}$, minus the one predicted by the DFT Hamiltonian diagonalization, $\rho^{elec}_{i,out}$. When computed in reciprocal space, the normal residual used in Broyden mixing is expressed as 
\begin{equation}
R^i(\bold{G})=\sum_{G\ne 0}\frac{1}{\bold{G}}\left[\rho^{elec}_{i,out}(\bold{G})-\rho^{elec}_{i,in}(\bold{G})\right],
\end{equation}
where $\bold{G}$ are reciprocal space lattice vectors and the sum is truncated according to the reciprocal space cutoff used for the plane-wave expansion of the electronic density. In the above equation, the total electronic charge of the system
\begin{equation}
\rho^{elec}(\bold{G}=\bold{0})=\int \rho^{elec}(\bold{r})e^{-i\bold{0}\cdot\bold{r}}d\bold{r}
\end{equation}
is not considered, as in standard SCF loops this quantity is not allowed to change. However, Sundararaman\cite{doi:10.1063/1.4978411} introduced a modification to the residual calculation that allows to also include the $(\bold{G}=0)$ component of the electronic charge density charge density 
\begin{equation}
R^i(\bold{G})=\sum_{G}\frac{1}{\bold{G}+\bold{G_c}}\left[\rho^{elec}_{i,out}(\bold{G})-\rho^{elec}_{i,in}(\bold{G})\right]
\end{equation}
Here, $\bold{G_c}$ is an empirically chosen parameter introduced to control the rate at which the total electronic charge is mixed during charge mixing. By using the above mixing method, the total charge density can be effectively adjusted during the self-consistent loop. This method was first implemented in the Quantum ESPRESSO (QE) code in combination with ESM\cite{Haruyama2018}. In this work, we extended the QE-ESM implementation of grand potential calculations to also allow the coupling with the Environ library of continuum embeddings and, in particular, with the SCCS model \cite{doi:10.1063/1.3676407} and its Poisson-Boltzmann extension for electrolyte solutions \cite{Nattino2019}. This new feature has been publicly released for Quantum ESPRESSO version 7.3 \cite{Giannozzi_2017} working with Environ version 3.1 \cite{Bainglass2022}. 

\subsection{Grand-Canonical Forces and Vibrational Properties}
The continuum embedding framework described above to handle solvent
and electrolyte effects allows us to perform self-consistent DFT calculations
with a preset electro-chemical potential, either directly, using a
fixed-potential scheme in the grand canonical ensemble, or indirectly,
by performing a Legendre transform on fixed-charge calculations in
the canonical ensemble. In the following, we review the relationships
between the critical properties of the system when computed using
the two approaches. 

We first discuss the total energy of the system, starting from its
expression in the canonical ensemble, $E^{C}(\alpha,N_{e})$, as a
function of the atomic degrees of freedom, $\alpha$, and the number
of electrons, $N_{e}.$ By Janak's theorem, 
\begin{equation}
\left(\frac{\partial E^{C}(\alpha,N_{e})}{\partial N_{e}}\right)_{\alpha}=E_{f}\equiv E_{f}^{*}(\alpha,N_{e})\label{eq:Janak}
\end{equation}
where we introduced the above notation to express the fact that, when
computed from simulations in the canonical ensemble, the Fermi energy,
$E_{f}$, becomes a function of the canonical natural variables. By
a Legendre transform, we can express the grand canonical energy as
\begin{equation}
E^{GC}(\alpha,E_{f})=E^{C}(\alpha,N_{e})-N_{e}E_{f},\label{eq:legendre}
\end{equation}
where now the Fermi energy is the natural state variable. The number
of electrons in this ensemble can be obtained from the partial derivative
of the grand canonical energy 
\begin{equation}
-\left(\frac{\partial E^{GC}(\alpha,E_{f})}{\partial E_{f}}\right)_{\alpha}=N_{e}\equiv N_{e}^{*}\left(\alpha,E_{f}\right)\label{eq:Janak2}
\end{equation}
and we adopt the same notation used above to indicate that, in grand
canonical simulations, the number of electrons becomes a function
of the ensemble's variables. 

Given that we can invert the Legendre transform, if we fix the Fermi
energy in a grand canonical calculation and use the corresponding
number of electrons in a canonical simulation, the resulting Fermi
energy corresponds to the initial one, i.e.
\begin{equation}
E_{f}^{*}(\alpha,N_{e}^{*}(\alpha,E_{f}))=E_{f}\label{eq:invert_ef}
\end{equation}
 and, vice versa,
\begin{equation}
N_{e}^{*}(\alpha,E_{f}^{*}(\alpha,N_{e}))=N_{e}.\label{eq:invert_ne}
\end{equation}
The above identities can be considered as an alternative definition
of the Legendre transform, for which the first derivative of the transformed
function is the inverse of the first derivative of the original function. 

The bidirectional relation between conjugated variables and the possibility
to express canonical and grand canonical energies using either of
the two introduce some unfortunate ambiguity in the practical application
of Eq. (\ref{eq:legendre}). By exploiting Eq. (\ref{eq:Janak2}),
it is possible to connect the grand canonical energy with its canonical
counterpart in terms of its natural variable, namely

\begin{equation}
E^{GC}(\alpha,E_{f})=E^{C}(\alpha,N_{e}^{*}(\alpha,E_{f}))-N_{e}^{*}(\alpha,E_{f})E_{f}.\label{eq:gc_wrt_c}
\end{equation}

Taking the partial derivative of the Eq. (\ref{eq:gc_wrt_c}) with
respect to one of the atomic coordinates allows us to derive the relationship
between interatomic forces in the two ensembles. Leveraging the fact
that the atomic displacements are not explicit functions of the conjugated
variables involved in the Legendre transform, we have
\begin{equation}
\left(\frac{\partial E^{GC}(\alpha,E_{f})}{\partial\alpha_{I\mu}}\right)_{E_{f}}=\left.\left(\frac{\partial E^{C}(\alpha,N_{e})}{\partial\alpha_{I\mu}}\right)_{N_{e}}\right|_{N_{e}^{*}}
\end{equation}
which implies that the forces in the grand canonical ensemble can
be expressed directly in terms of the ones from the canonical ensemble,
and we introduced a notation to specify that the partial derivative
is evaluated for a number of electrons that correspond to the value
obtained in the grand canonical simulation under a potential $E_{f}$,
\begin{equation}
f_{I\mu}^{GC}(\alpha,E_{f})=f_{I\mu}^{C}(\alpha,N_{e}^{*}(\alpha,E_{f})).\label{eq:fgc_equal_fc}
\end{equation}

Following the above derivation, we can investigate the relationship
between the second derivatives of the energy with respect to atomic
displacements in the two ensembles. The matrix of force constants
in the grand canonical ensemble is given by 
\begin{equation}
c_{IJ\mu\nu}^{GC}\left(\alpha,E_{f}\right)=\left(\frac{\partial}{\partial\alpha_{J\nu}}\frac{\partial}{\partial\alpha_{I\mu}}E^{GC}\left(\alpha,E_{f}\right)\right)_{E_{f}}=\left(-\frac{\partial}{\partial\alpha_{J\nu}}f_{I\mu}^{GC}\left(\alpha,E_{f}\right)\right)_{E_{f}},
\end{equation}
where $I$and $J$ are associated with the atomic indexes, while $\mu$and
$\nu$ indicate the Cartesian directions (x, y, and z). Exploiting
the result in Eq. (\ref{eq:fgc_equal_fc}) we can show that

\begin{equation}
c_{IJ\mu\nu}^{GC}\left(\alpha,E_{f}\right)=\left.\left(\frac{\partial^{2}E^{C}\left(\alpha,N_{e}\right)}{\partial\alpha_{J\nu}\partial\alpha_{J\nu}}\right)_{N_{e}}\right|_{N_{e}^{*}}+
\left.\left(\frac{\partial N_{e}^{*}\left(\alpha,E_{f}\right)}{\partial\alpha_{J\nu}}\right)_{E_{f}}\right|_{E_{f}^{*}}\cdot
\left.\left(\frac{\partial E_{f}^{*}\left(\alpha,N_{e}\right)}{\partial\alpha_{I\mu}}\right)\right|_{N_{e}^{*}}\label{eq:cmatrix_gc_wrt_c}
\end{equation}
The first term on the right-hand side of Eq. (\ref{eq:cmatrix_gc_wrt_c})
corresponds to the canonical force constants, while the second term
involves the derivative of the Legendre product between conjugated
variables. By exploiting Eqs.(\ref{eq:invert_ef}) and (\ref{eq:invert_ne}),
we can rewrite Eq. (\ref{eq:cmatrix_gc_wrt_c}) in a more symmetric
form as
\begin{eqnarray}
c_{IJ\mu\nu}^{GC}\left(\alpha,E_{f}\right) & = & c_{IJ\mu\nu}^{C}\left(\alpha,N_{e}^{*}\left(\alpha,E_{f}\right)\right)-\nonumber \\
 &  & \left.\left(\frac{\partial N_{e}^{*}\left(\alpha,E_{f}\right)}{\partial E_{f}}\right)_{\alpha}\right|_{N_{e}^{*}}
 \cdot\left.\left(\frac{\partial E_{f}^{*}\left(\alpha,N_{e}\right)}{\partial\alpha_{J\nu}}\right)_{N_{e}}\right|_{N_{e}^{*}}\cdot\left.\left(\frac{\partial E_{f}^{*}\left(\alpha,N_{e}\right)}{\partial\alpha_{I\mu}}\right)_{N_{e}}\right|_{N_{e}^{*}}\label{eq:cmatrix_gc_ef}\\
c_{IJ\mu\nu}^{GC}\left(\alpha,E_{f}\right) & = & c_{IJ\mu\nu}^{C}\left(\alpha,N_{e}^{*}\left(\alpha,E_{f}\right)\right)-\nonumber \\
 &  & \left.\left(\frac{\partial E_{f}^{*}\left(\alpha,N_{e}\right)}{\partial N_{e}}\right)_{\alpha}\right|_{E_{f}^{*}}
 \cdot\left.\left(\frac{\partial N_{e}^{*}\left(\alpha,E_{f}\right)}{\partial\alpha_{J\nu}}\right)_{E_{f}}
 \right|_{E_{f}^{*}}\cdot
 \left.\left(\frac{\partial N_{e}^{*}\left(\alpha,E_{f}\right)}{\partial\alpha_{I\mu}}\right)_{E_{f}}\right|_{E_{f}^{*}}\label{eq:cmatrix_gc_ne}
\end{eqnarray}

Eq. (\ref{eq:cmatrix_gc_wrt_c}) reveals that, under fixed-potential conditions, the force constant matrix—and consequently the vibrational frequencies—can differ from those computed under fixed-charge conditions. The origin of this difference lies in the variation of the electron number as atoms vibrate in grand canonical condition, which modifies the system’s energy curvature with respect to atomic displacements. An equivalent relation has also been reported in \cite{GrandCanonicalHormann}, consistent with the result we derived independently. Moreover, as shown in Eqs. (\ref{eq:cmatrix_gc_ne}) and (\ref{eq:cmatrix_gc_ef}), the force constants under one condition (canonical or grand canonical) can be derived from calculations performed under the other, providing a practical bridge between the two thermodynamic frameworks.

From a practical standpoint, directly evaluating the correction term remains challenging. While continuum embedding methods have been combined with linear response density functional perturbation theory for electronic excitations \cite{Timrov2015}, no comparable implementation exists for atomic displacements. Therefore, in the following, we employ a finite-difference approach to compute force constant matrices in both canonical and grand canonical ensembles.

\subsection{Surface Capacitance and Chemical Hardness}
\label{sec:surface_capacitance}
Eqs.(\ref{eq:cmatrix_gc_ne}) and (\ref{eq:cmatrix_gc_ef}) involve the changes in the chemical potential
or number of electrons as one atom of the system is moved, while keeping
the conjugated variable fixed. The factor multiplying these terms
can be associated with either the differential capacitance 
\begin{equation}
C\left(E_{f}\right)=\left(\frac{\partial N_{e}^{*}\left(\alpha,E_{f}\right)}{\partial E_{f}}\right)_{\alpha}
\label{surface capacitance}
\end{equation}
or the chemical hardness of the electrified interface
\begin{equation}
\eta\left(E_{f}\right)=\left.\left(\frac{\partial E_{f}^{*}\left(\alpha,N_{e}\right)}{\partial N_{e}}\right)_{\alpha}\right|_{N_{e}^{*}}=\frac{1}{C\left(E_{f}\right)}.
\end{equation}
Leveraging the derivation of Binninger \cite{Binninger2021},
we can express the chemical hardness of an electrified interface in the presence of a continuum description of the electrolyte (dielectric screening and ionic countercharge) as the sum of three terms
\begin{equation}
\eta\left(E_{f}\right)=\eta_{Q}\left(E_{f}\right)+\eta_{XC}\left(E_{f}\right)+\eta_{DL}\left(E_{f}\right).
\end{equation}
The quantum hardness is related to the finite-temperature density
of states of the electronic system, $g_{D}^{T}\left(E_{f}\right)$,
as
\[
\eta_{Q}\left(E_{f}\right)=\frac{1}{g_{D}^{T}\left(E_{f}\right)},
\]

The exchange-correlation hardness can be expressed as
\begin{equation}
\eta_{XC}=\int\frac{g_{LD}^{T}\left(E_{f},\mathbf{r}\right)}{g_{D}^{T}\left(E_{f}\right)}\left(\int\frac{\delta\mu_{XC}\left(\mathbf{r}\right)}{\delta\rho^{elec}\left(\mathbf{r}'\right)}f^{elec}\left(\mathbf{r}'\right)d\mathbf{r}'\right)d\mathbf{r},
\end{equation}
where $\mu_{XC}\left(\mathbf{r}\right)$ is the exchange-correlation
functional used in the electronic structure calculation, $g_{LD}^{T}\left(E_{f},\mathbf{r}\right)$
is the local density of states at the Fermi energy, and 
\begin{equation}
f^{elec}\left(\mathbf{r}\right)=\frac{\partial\rho^{elec}\left(\mathbf{r}\right)}{\partial N},
\end{equation}
is the electronic Fukui function of the system. 

Eventually, the electrostatic component of the chemical hardness, coming from the electrostatic potential due to the electrons and the continuum diffuse layer charge reorganization, is expressed as
\begin{equation}
\eta_{DL}=\int\frac{g_{LD}^{T}\left(E_{f},\mathbf{r}\right)}{g_{D}^{T}\left(E_{f}\right)}\left(\phi^{f}\left(\mathbf{r}\right)-\phi_{b}^{f}\right)d\mathbf{r},
\end{equation}
with 
\[
\phi^{f}=\int\frac{f^{elec}\left(\mathbf{r}'\right)-f^{DL}\left(\mathbf{r}'\right)}{\left|\mathbf{r}-\mathbf{r}'\right|}d\mathbf{r}'
\]
 being the electrostatic potential generated by the electronic and diffuse-layer Fukui functions, with an asymptotic value of $\phi_{b}^{f}$ in the bulk of the electrolyte. Similar to the electronic Fukui function, the diffuse-layer counterpart is defined as
\begin{equation}
f^{DL}\left(\mathbf{r}\right)=\frac{\partial\rho^{diel}\left(\mathbf{r}\right)}{\partial N}+\frac{\partial\rho^{ion}\left(\mathbf{r}\right)}{\partial N}'.
\end{equation}
It allows to accounts for the flow in charge in the environment as the charge of the substrate is changed, as imposed by the requirement of charge neutrality in the simulation cell, with
\begin{equation}
\int f^{elec}\left(\mathbf{r}\right)d\mathbf{r}=1\quad\text{and}\quad\int f^{DL}\left(\mathbf{r}\right)d\mathbf{r}=1.
\end{equation}
When the metal is sufficiently thick, the density of states at the Fermi energy is large enough, so that the quantum hardness negligible. Moreover, since the electronic Fukui function is localized near the metal electrode surface, and the normalized local density of states (LDOS) is periodic within the bulk of the electrode, the weight part $\frac{g_{LD}^{T}\left(E_{f},\mathbf{r}\right)}{g_{D}^{T}\left(E_{f}\right)}$ of the bulk region approaches 1 for thick electrodes. So, the exchange-correlation hardness also becomes negligible, and the diffuse layer term converges to the potential drop, $\Delta\phi^{f}$, induced by the electrochemical Fukui densities:
\begin{equation}
\eta_{DL}=\frac{1}{C_{DL}}=\Delta\phi^{f}\propto\frac{1}{A}\int z\left(f^{elec}\left(\mathbf{r}'\right)-f^{DL}\left(\mathbf{r}'\right)\right)d\mathbf{r}
\label{surface area}
\end{equation}
where $A$ is the surface area of the slab. This term is proportional to the average surface dipole generated by these densities. Therefore, for a sufficiently thick metal slab, the chemical hardness is inversely proportional to the surface area, while the surface capacitance is directly proportional to the surface area.

Moreover, the change in Fermi energy with respect to atomic displacements, $\left.\left(\frac{\partial E_{f}(\alpha,N_{e})}{\partial \alpha_{I\mu}}\right)\right|{N{e}}$, is inversely proportional to the surface area. In contrast, the change in the number of surface electrons induced by atomic displacements does not scale with the surface area. This is because, in canonical ensemble as atom move at fixed N, the induced work function change scales as 1/A. Then, the number of electrons to add or remove to compensate for that change will be proportional to the capacitance times Delta Phi thus scaling as A times 1/A, and thus independent of the surface area. Given these scaling behaviors, it follows from Eqs.~(\ref{eq:cmatrix_gc_ef}) and (\ref{eq:cmatrix_gc_ne}) that the difference in force constants between the canonical and grand-canonical conditions scales inversely with the surface area. Since vibrational frequencies depend on the square root of the force constants, the change in frequency is proportional to the change in force constants. In our case, the variation in force constants is small compared to their absolute values, allowing the frequency shift to be approximated as linearly proportional to the force constant variation. Therefore, the frequency difference between canonical and grand-canonical conditions is also expected to scale approximately inversely with the surface area.

\section{Computational Details}

\begin{figure}
\centering
\subfigure[Dielectric boundary]{
\label{solvation boundary}
\includegraphics[width=0.25\textwidth]{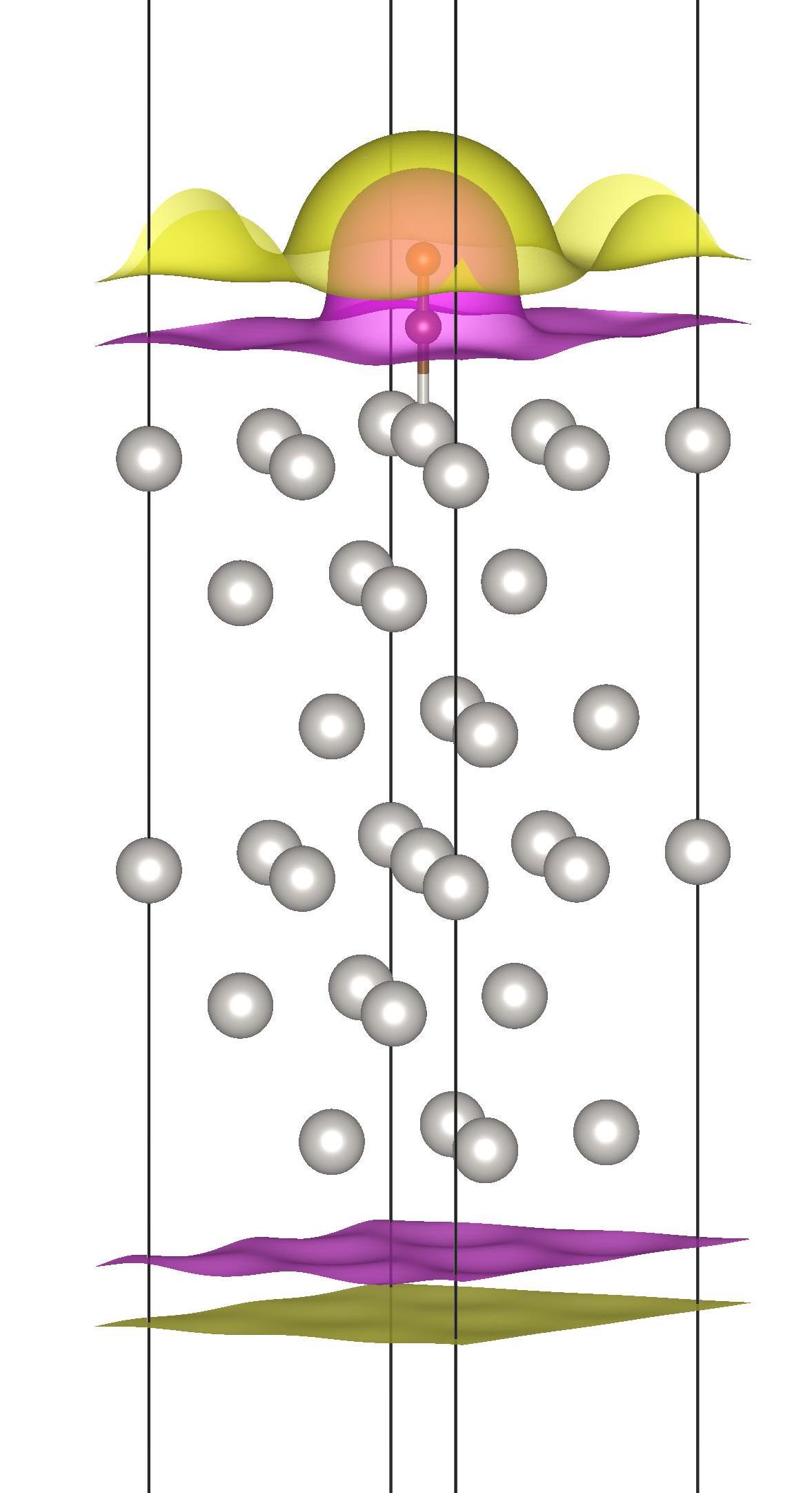}
}
\subfigure[Electrolyte boundary]{
\label{electrolyte boundary}
\includegraphics[width=0.25\textwidth]{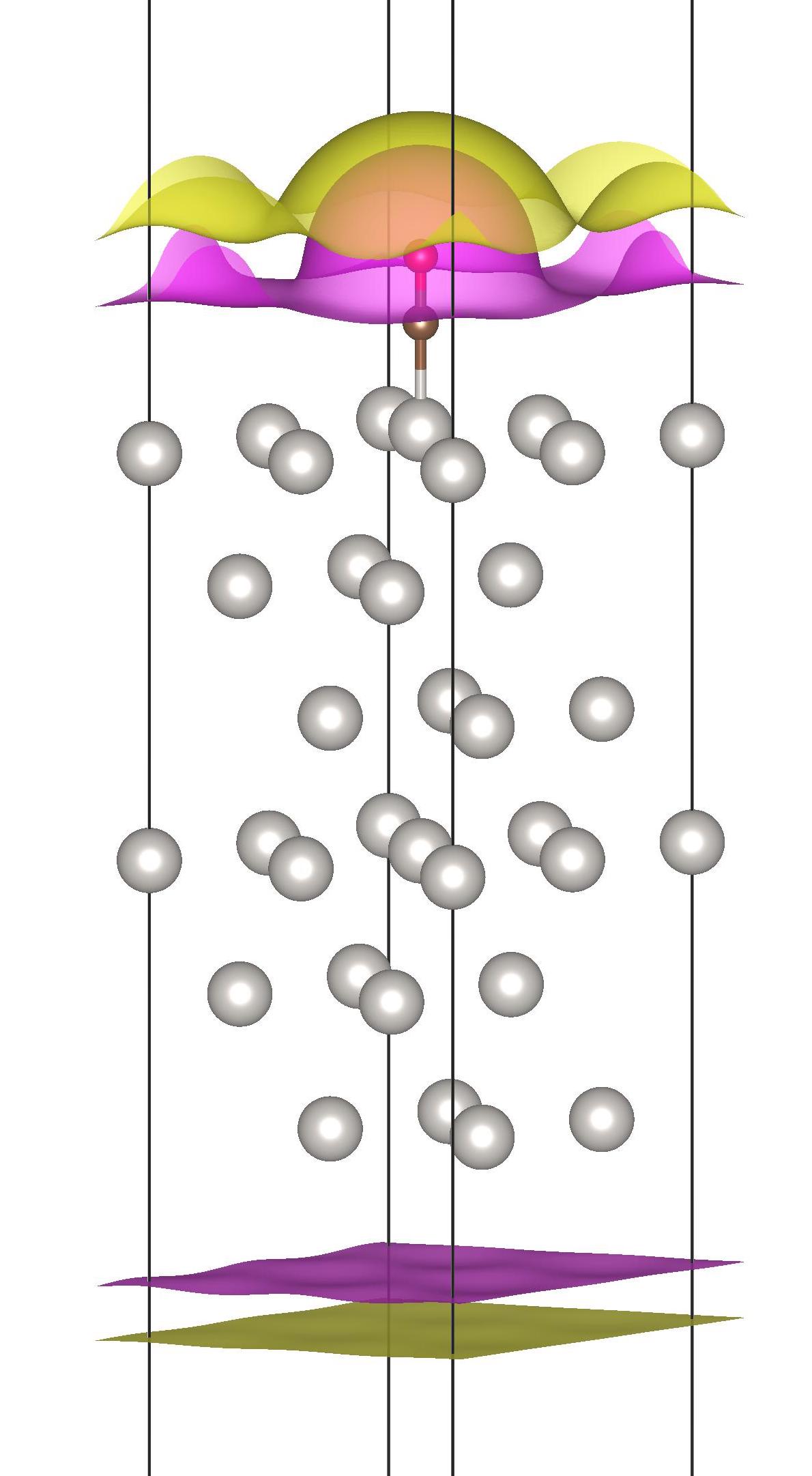}}
\caption{Atomic structure of the selected test systems (round spheres) together with the continuum interfaces adopted for the solvent, in panel (a), and the electrolyte, in panel (b). For the solvent's continuum interface (left), the reported yellow and purple iso-surfaces correspond to $\rho_{min,\epsilon}=1.2\times 10^{-4}$a.u. and  $\rho_{max,\epsilon}=2.2\times 10^{-3}$a.u., lying approximately at a distance from the Pt surface of 2.0$\AA$ and 3.1$\AA$, respectively. For the electrolyte's continuum interface, the yellow and violet iso-surfaces correspond to $\rho_{min,\gamma}=3\times 10^{-5}$a.u. and  $\rho_{max,\gamma}=4\times 10^{-4}$ a.u., lying approximately at a distance from the Pt surface of 2.6$\AA$ and 3.8$\AA$, respectively. }
\label{boundary figure} 
\end{figure}

We consider carbon monoxide adsorbed on a Pt(111) surface as the testing example for the presented derivations. As the CO molecule adsorbed on Pt surfaces has been extensively studied as a model system for understanding chemisorption and vibrational behavior at electrochemical interfaces. Its well-characterized vibrational response to the applied potential—known as the Stark tuning rate—makes it a reliable probe of the interfacial electrostatic environment. In particular, the C–O stretching frequency is sensitive to changes in surface environments, which are key electrochemical descriptors of electrode surfaces under bias.

The whole calculation is performed in Quantum ESPRESSO code\cite{Giannozzi_2017, Giannozzi_2009} with Quantuim ENVIRON library\cite{doi:10.1063/1.3676407}. Norm conserving SG15 pseudopotential\cite{Hamann1979, Hamann2013, Schlipf2015} is used in all our calculations. To ensure the accuracy of atomic forces and vibrational frequencies obtained from finite-difference calculations, we conducted convergence tests with respect to three key computational parameters: the plane-wave energy cutoff, the vacuum layer thickness, and the metal slab thickness. The results show that, with appropriately chosen parameters, the numerical errors in atomic forces can be reduced to within $2meV/Å$, and the errors in vibrational frequencies can be kept below $2cm^{-1}$.  In addition, we examined the consistency between Hellmann–Feynman forces and finite-difference forces in the presence of a dielectric environment and the linearized Poisson–Boltzmann effect. The difference between the two approaches is found to be less than $0.02meV/Å$, confirming the reliability of Hellmann–Feynman force calculations under these conditions. A detailed description of these convergence tests is provided in the Supporting Information.

The atomic structure, solvent boundary, and electrolyte boundary used in the simulations are visualized in Figure~\ref{boundary figure}. In practice, this required us to adjust the values of the two parameters used to define the position of the continuum interfaces for both the solvent (Equation~\eqref{frho}) and the electrolyte (the $\gamma(\bold{r})$ function introduced in Eq~\eqref{gamma})\cite{Dabo2008_stark}. In the simulations presented in the following sections, the boundary of the solvent dielectric is set by the parameters of $\rho_{min,\epsilon}=1.2\times 10^{-5}$ and  $\rho_{max,\epsilon}=2.2\times 10^{-4}$, as shown in Figure~\ref{solvation boundary}. Similarly, the boundary for the electrolyte is set to adopt  $\rho_{min,\gamma}=3\times 10^{-6}$ and  $\rho_{max,\gamma}=4\times 10^{-5}$, as shown in Figure~\ref{electrolyte boundary}.

Vibrational frequencies are computed using a finite difference approach implemented within the AiiDA ENVIRON workflow \cite{Truscott2021} as part of the AiiDA framework\cite{Huber2020, UHRIN2021110086}. To simplify and reduce the costs of the vibrational calculations, only the carbon and oxygen atoms were allowed to move when constructing the matrix of force constants. Owing to the large mass of the Pt atoms, this approximation yields vibrational frequencies that agree with those obtained from a full dynamical matrix calculation involving the entire system \cite{BONNET2014210}. In order to compare results in the two considered ensembles, calculations were first performed in the canonical ensemble using a fixed number of electrons. Fixed-potential grand-canonical calculations were then carried out by setting the Fermi energy to the value obtained from the canonical ensemble, allowing for a more direct comparison of vibrational frequencies at the same potential.

\section{Results and Discussion}
\subsection{Finite-Difference Calculation of Stark Tuning Rates for CO on the Pt(111) Surface}

As discussed in the previous section, we expect the difference in vibrational frequencies between the two ensembles to originate from charge redistribution during atomic displacements. This effect manifests as electron number variation in the grand-canonical ensemble, and as Fermi level variation in the canonical ensemble. To verify this reasoning, we consider the direction of the atomic vibrations. When atoms vibrate parallel to the surface, the distance between the molecule and the metal slab remains nearly unchanged, leading to negligible variations in surface charge. In contrast, perpendicular vibrations alter this distance significantly, resulting in a notable change in surface charge. Accordingly, we compare the vibrational frequencies along directions parallel and perpendicular to the surface under both grand-canonical and canonical ensemble. 

Furthermore, in the limit of an infinitely large surface, the surface charge remains effectively constant during atomic displacements in the perpendicular direction, causing the vibrational frequency difference between the ensembles to vanish. To examine this size dependence, we computed the vibrational frequencies and Stark tuning rates of CO on Pt(111) surfaces modeled with increasing lateral dimensions: 2×2, 3×3, and 4×4 unit cells. In our setup, each unit cell contains one CO molecule, therefore, with increasing surface area, the simulations effectively probe different coverage rates. This setup enables us to assess how both coverage and surface area affect vibrational properties, providing a consistent test of the expected scaling behavior.

In addition, we quantitatively validate Eq.(\ref{eq:cmatrix_gc_ef}) by comparing its predicted grand-canonical vibrational frequencies with those obtained from explicit grand-canonical finite-difference calculations. This comparison also confirms that, by applying Eq.(\ref{eq:cmatrix_gc_ef}) to canonical finite-difference results, one can directly obtain the corresponding grand-canonical frequencies without performing separate finite-difference calculations under grand-canonical ensemble.

\begin{figure}
\centering
\subfigure[vib. in XY plane for 1/4 CO coverage]{
\includegraphics[width=0.4\textwidth]{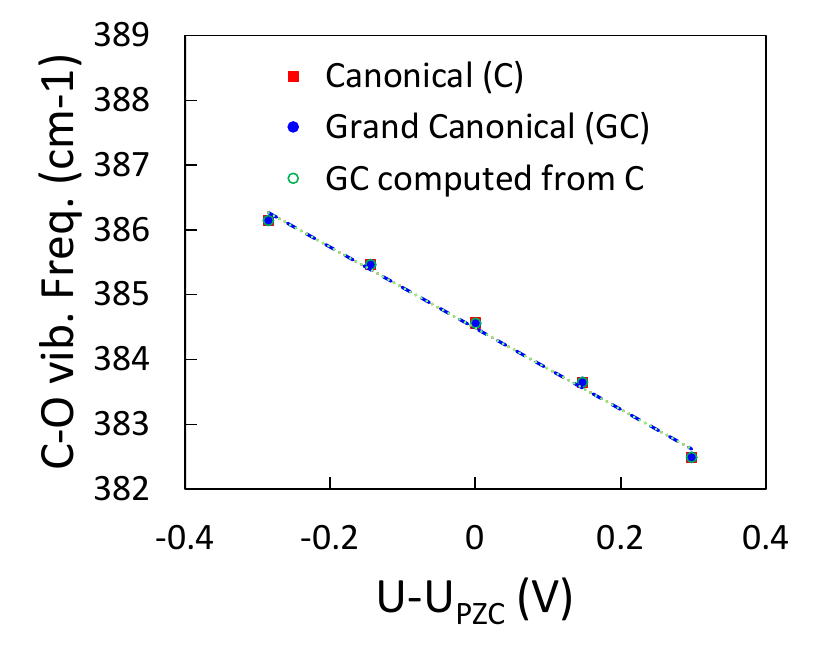}
\label{Stark rate on xy plane}
}
\subfigure[vib. in Z direction for 1/4 CO coverage]{
\includegraphics[width=0.4\textwidth]{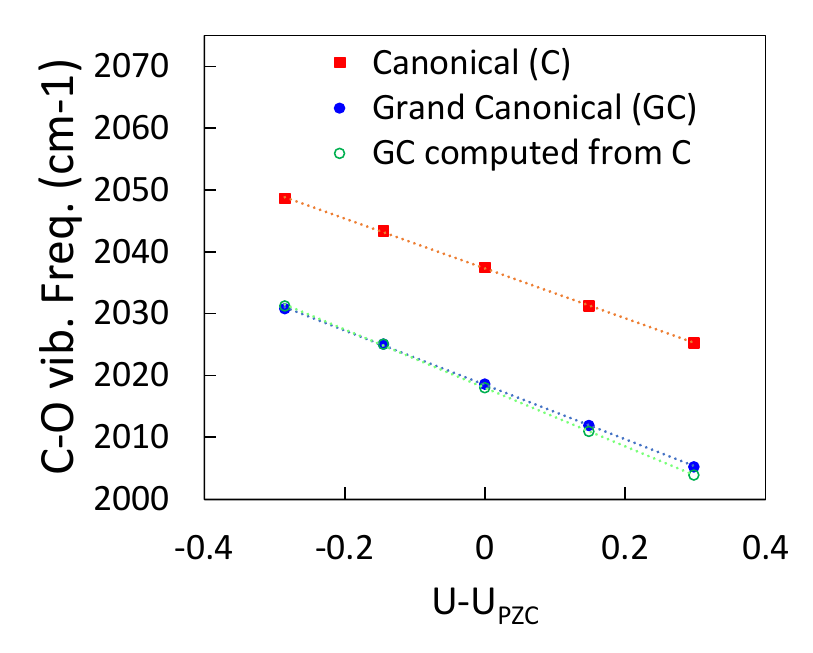}
\label{Stark rate on 1/4 coverage}
}
\subfigure[vib. in Z direction for 1/9 CO coverage]{
\includegraphics[width=0.4\textwidth]{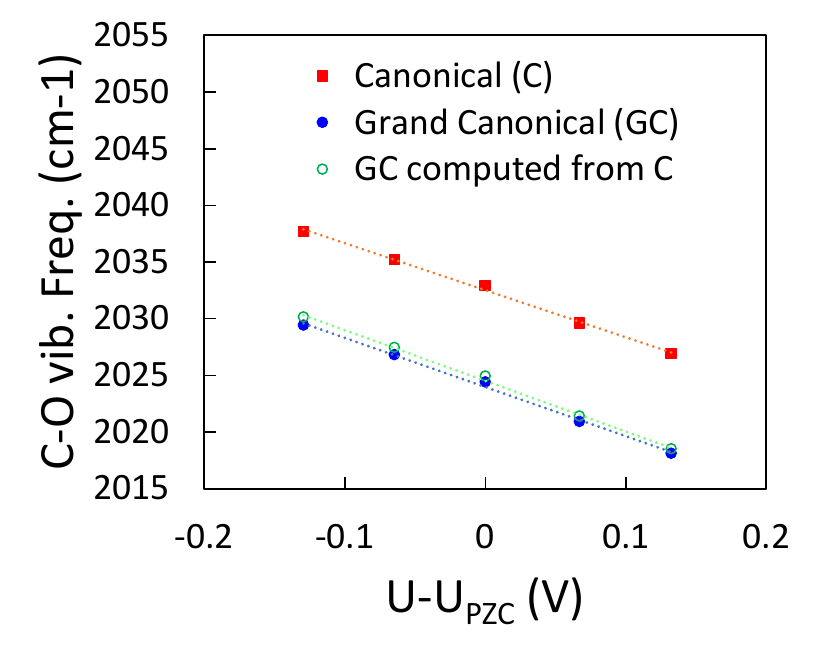}
\label{Stark rate on 1/9 coverage}
}
\subfigure[vib. in Z direction for 1/16 CO coverage]{
\includegraphics[width=0.4\textwidth]{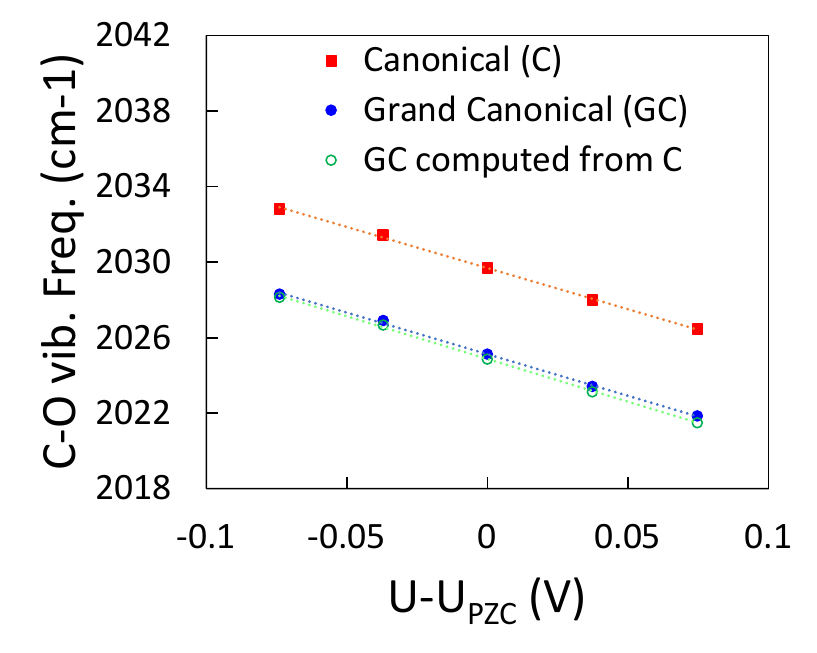}
\label{Stark rate on 1/16 coverage}
}
\caption{ Vibrational frequencies of selected modes as a function of the Fermi energy.
(a) Modes in the xy plane (parallel) for 1/4 CO coverage;
(b) Modes in the z direction (perpendicular) for 1/4 CO coverage;
(c) z-direction modes for 1/9 CO coverage;
(d) z-direction modes for 1/16 CO coverage.
Frequencies computed under grand-canonical and canonical ensemble using finite-difference methods are shown in blue and red colors respectively.
Grand-canonical frequencies computed from canonical data using 
are shown in green symbols.
Linear fits are used to extract Stark tuning rates from all three data sources.
}
\label{Stark rate}
\end{figure}

Figure~\ref{Stark rate} summarizes the calculated vibrational frequencies for the CO vibrations parallel (in panel (a)) and perpendicular (in panel (b)) to the Pt surface using a 2x2 Pt supercell. The effect of system size is also reported in Figure~\ref{Stark rate}, where results of the perpendicular vibrational model computed for the 3x3 (in panel (c)) and 4x4 (in panel (d)) supercells are reported. In each panel, the dependence of vibrational frequencies on the applied voltage (i.e., Fermi level) is indicated by color: the grand canonical results are shown in blue, while the canonical results are shown in red.

\begin{figure}
\centering
\subfigure[]{
\includegraphics[width=0.4\textwidth]{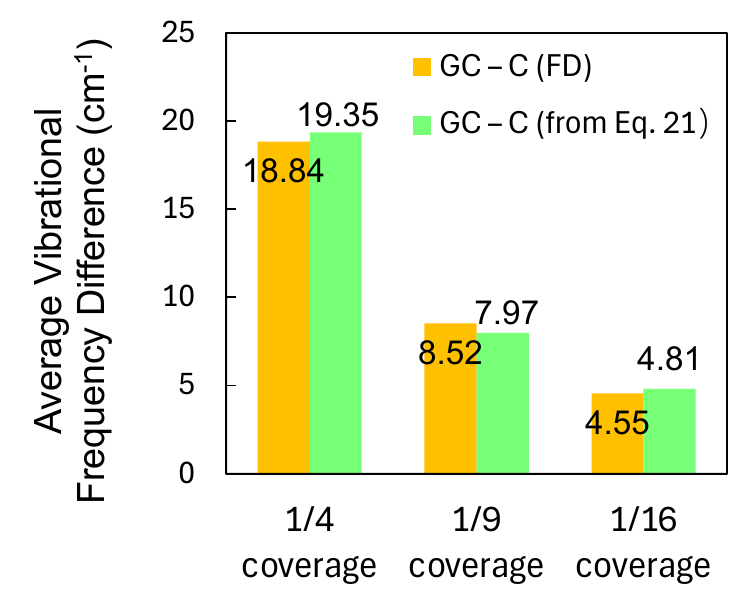}
\label{Comparison freq ave diff}
}
\subfigure[]{
\includegraphics[width=0.4\textwidth]{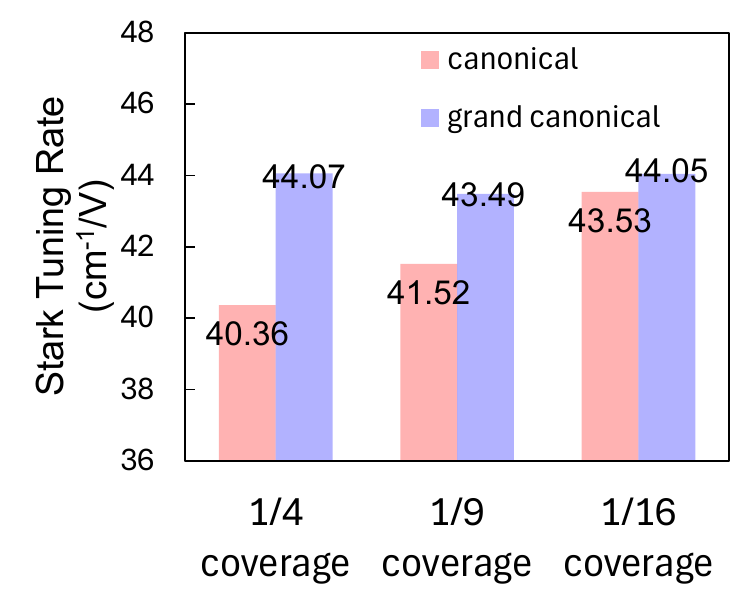}
\label{Comparison stark rate}
}
\caption{Comparison of vibrational properties computed under canonical and grand-canonical ensemble for Pt(111) surfaces of different coverage (1/4, 1/9, and 1/16).
(a) shows the average vibrational frequency differences between the two ensembles. Orange bars represent values directly calculated via finite-difference (FD) methods, while green bars are obtained by applying Eq. (\ref{eq:cmatrix_gc_ef}) to canonical results to compute grand-canonical frequencies.
(b)  presents the Stark tuning rates, extracted from the slopes of vibrational frequencies as a function of the Fermi energy, under both canonical (red bars) and grand-canonical (blue bars) conditions.}
\label{Comparison}
\end{figure}

The difference in the frequency of the parallel vibrational mode between canonical and grand canonical simulations is rather small and of the order of $0.1cm^{-1}$ for all the electro-chemical potentials considered. The Stark tuning rate also shows very close agreement between the two approaches. However, as shown in Figure~\ref{Comparison freq ave diff}, the perpendicular vibrational frequency under canonical conditions is on average $18.8cm^{-1}$ higher than under grand-canonical conditions. The different ensembles also affect the Stark tuning rate of the perpendicular mode, with a difference of $3.7cm^{-1}/eV$ shown as the difference between the red bar and the blue bar in Figure~\ref{Comparison stark rate}. 

To evaluate the effect of surface size, we observe that the vibrational frequency difference decreases as the surface size increases, going from Figure~\ref{Stark rate on 1/4 coverage} to Figure~\ref{Stark rate on 1/9 coverage} and then to Figure~\ref{Stark rate on 1/16 coverage}. The average frequency difference decreases from approximately $18.8cm^{-1}$ to $7.36cm^{-1}$ and further to $4.55cm^{-1}$. This trend is consistent with the theoretical analysis presented in Section~\ref{sec:surface_capacitance}, which predicts that the change in vibrational frequencies due to atomic displacements scales inversely with the surface area, or equivalently, with the surface occupation rate in our setup.

We can also find the same trend on Stark tuning rate, as shown in Figure~\ref{Comparison stark rate}, as the difference in Stark tuning rate between grand-canonical and canonical boundary is very small if atom vibrates parallel to metal surface. On the other hand, the difference is significant for the vibrational model is perpendicular to metal surface, and the difference decreases as the surface area increases, or the occupation rate decrease. 

In the validation of Eq.(\ref{eq:cmatrix_gc_ef}), we computed both $\left.\left(\frac{\partial N_{e}^{}\left(\alpha,E_{f}\right)}{\partial\alpha_{I\mu}}\right){E{f}}\right|{E{f}^{}}$ and the surface capacitance term $\left.\left(\frac{\partial E_{f}^{}\left(\alpha,N_{e}\right)}{\partial N_{e}}\right){\alpha}\right|{E_{f}^{}}$ using finite-difference calculations under canonical ensemble. By combining these two quantities through Eq.(\ref{eq:cmatrix_gc_ef}), we obtained the correction to the force constant matrix that accounts for grand-canonical ensemble. This corrected force constant matrix yields a modified dynamical matrix, from which the grand-canonical vibrational frequencies can be computed.

The resulting frequencies are plotted as green markers in Fig~(\ref{Stark rate}). These predicted grand-canonical frequencies show excellent agreement with those obtained from explicit grand-canonical finite-difference calculations (blue markers). In Fig~(\ref{Comparison freq ave diff}), green bars represent the frequency differences between canonical and predicted grand-canonical results using Eq.(\ref{eq:cmatrix_gc_ef}), while orange bars correspond to the differences obtained from direct grand-canonical finite-difference calculations. The close match between the two confirms the validity of our analytical formula, Eq.(\ref{eq:cmatrix_gc_ef}), with an average discrepancy of less than $1,\mathrm{cm}^{-1}$. This strong agreement demonstrates that vibrational frequencies under grand-canonical and canonical ensembles do exhibit systematic differences, and that these differences can be captured by the correction terms given in Eq.(\ref{eq:cmatrix_gc_ef}) and Eq.~(\ref{eq:cmatrix_gc_ne}).

\subsection{Effects of Implicit Solvent Parameters on Stark Tuning Rates}

One may have noticed that the experimental stark tuning rate is around $29cm^{-1}/V$\cite{Lozovoi2007, Luo1993, Lin2000}, whereas the grand-potential stark tuning rate  obtained from our calculation is $41.0cm^{-1}/V$, which is larger than the experimental result, and even worse than that in canonical ensemble result of $35.5cm^{-1}/V$. This discrepancy can be attributed to two main factors.

First, it relates to the ordered structure of interfacial water molecules at the metal surface. Previous molecular dynamics studies have shown that the dielectric constant of surface-confined water in the z-direction ranges between 4 and 9 \cite{Dinpajooh2016, Palinkas1977}, significantly lower than the bulk value of 78. However, in continuum models of implicit solvent, the dielectric constant is typically assumed to be a spatially uniform constant of 78. This overestimation of screening likely leads to an exaggerated field response. Therefore, a smaller and spatially varying dielectric constant should be considered to better capture the local electrostatic environment at the metal–solution interface.

\begin{figure}
\centering
\subfigure[$\epsilon_0=6.0$]{
\includegraphics[width=0.3\textwidth]{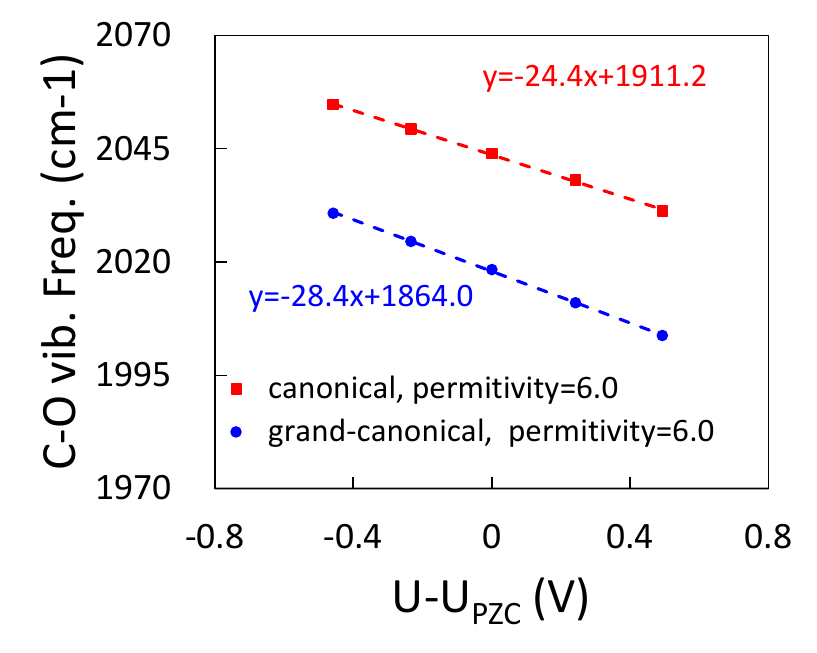}
\label{Stark rate with permittivity of 6}
}
\subfigure[Reduced interface distances]{
\includegraphics[width=0.3\textwidth]{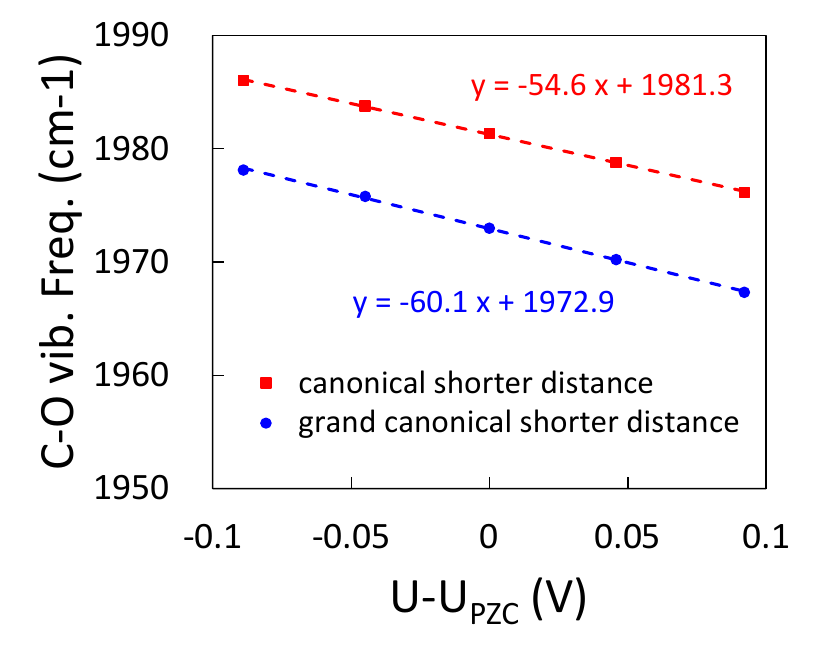}
\label{Stark rate with shorter distance}
}
\subfigure[Increased interface distances]{
\includegraphics[width=0.3\textwidth]{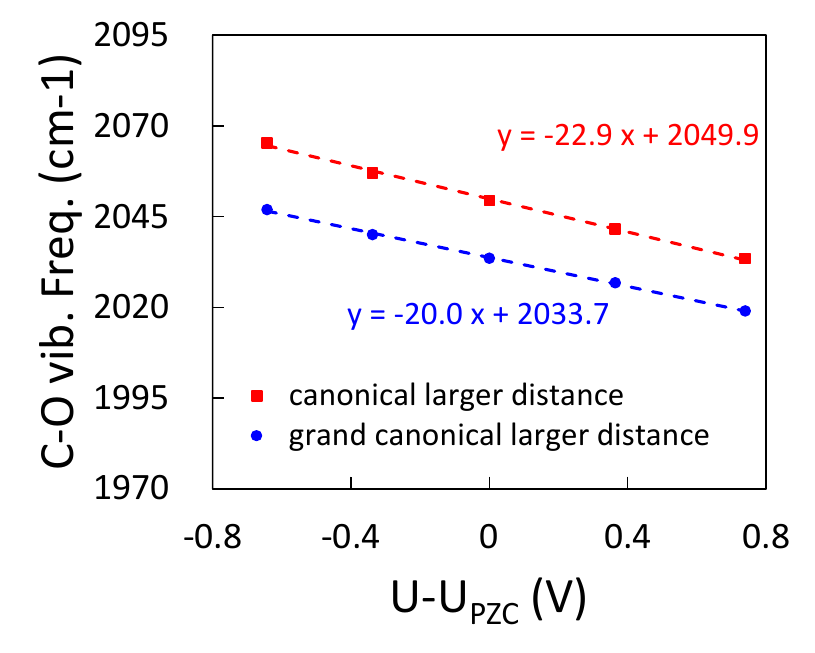}
\label{Stark rate with longer distance}
}
\caption{Comparison of Stark tuning rates calculated by different solvation and electrolyte environment parameters. (a) Stark tuning rates with solvation permittivity =6. (b) Stark tuning rates with shorter solvation and electrolyte interface distance. (c) Stark tuning rates with longer solvation and electrolyte interface distance.}
\label{Stark rate with diff sol parameters}
\end{figure}

To better reflect the ordered nature of surface water, we performed additional calculations using a dielectric constant of 6, which lies within the reported range. The results, shown in Figure~\ref{Stark rate with permittivity of 6}, demonstrate that the grand-potential Stark tuning rate and canonical Stark tuning rate are $28.4cm^{-1}/V$ and $24.4cm^{-1}/V$ respectively. These values are significantly lower than those calculated using a dielectric constant of 78 and are much closer to the experimental result. Moreover, the gap between the grand-potential and canonical conditions becomes notably smaller.

Second, the discrepancy also arises from how implicit models define the interface between the metal surface and the surrounding solvent and electrolyte. These models treat the solvent and electrolyte as statistical averages of their respective phases, and their interactions with the metal surface and the adsorbed CO molecule are governed by mean-field electrostatics. The strength and spatial distribution of these interactions depend critically on parameters describing the electrolyte distribution and solvent–metal interface. Previous studies have typically calibrated these parameters using solvation energies, the work function in solvent, or surface capacitance. \cite{doi:10.1063/1.3676407, doi:10.1063/1.5054580} These findings underscore a critical feature of implicit solvent models: the predicted physical properties are highly sensitive to multiple adjustable parameters. Since detailed parameter optimization is beyond the scope of this work, we instead demonstrate the sensitivity of the calculated tuning rate by adjusting the interface distances—once by increasing and once by decreasing both solvent and electrolyte distances simultaneously. The resulting variation highlights the critical importance of proper parameter selection and calibration in implicit electrochemical models.

In the first case, we used smaller interface distances: $\rho_{\text{min},\epsilon} = 1.2 \times 10^{-3}$ a.u. and $\rho_{\text{max},\epsilon} = 2.2 \times 10^{-2}$ a.u. for the solvent, and $\rho_{\text{min},\gamma} = 3 \times 10^{-4}$ a.u. and $\rho_{\text{max},\gamma} = 4 \times 10^{-3}$ a.u. for the electrolyte. These choices led to a much higher surface capacitance of $34.5~\mu\text{F/cm}^2$, and the corresponding Stark tuning rates, as shown in Figure~\ref{Stark rate with shorter distance},  were $60.1~\text{cm}^{-1}/\text{V}$ under grand canonical ensemble and $54.6~\text{cm}^{-1}/\text{V}$ under canonical ensemble.

In the second case, we used relatively larger distances between the metal and both the implicit solvent and electrolyte regions. Specifically, for the solvent continuum interface, we set $\rho_{\text{min},\epsilon} = 3.0 \times 10^{-6}$ a.u. and $\rho_{\text{max},\epsilon} = 5.5 \times 10^{-5}$ a.u.; for the electrolyte continuum interface, we used $\rho_{\text{min},\gamma} = 5 \times 10^{-7}$ a.u. and $\rho_{\text{max},\gamma} = 5 \times 10^{-6}$ a.u. Under these settings, the computed surface capacitance was $4.5~\mu\text{F/cm}^2$. The resulting Stark tuning rates, shown in Figure~\ref{Stark rate with longer distance}, were $22.9~\text{cm}^{-1}/\text{V}$ (grand canonical) and $20.0~\text{cm}^{-1}/\text{V}$ (canonical). 

We observe that the computed Stark tuning rates are not necessarily more accurate when the surface capacitance is closer to the experimental value (approximately $40~\mu\text{F/cm}^2$). In fact, configurations with surface capacitances significantly lower than the experimental range (e.g., achieved by using larger interface distances) sometimes yield Stark tuning rates that better match experimental data. This counterintuitive trend highlights the complexity of parameter selection and calibration in implicit electrochemical models.

These findings underscore a critical feature of implicit solvent models: the predicted physical properties are highly sensitive to multiple adjustable parameters, including the dielectric constant, the solvent–metal interface positioning, and the electrolyte profile. The interdependence of these parameters making careful parameter selection important for targeting different experimental observables. To overcome these issues, future developments may involve hybrid approaches that introduce explicit interfacial water layers while retaining implicit bulk behavior, or models that allow for a spatially varying dielectric response. However, such improvements lie beyond the scope of this work. Here, we focus on analyzing how interfacial model parameters affect the Stark tuning rate and comparing results under grand-canonical and canonical ensembles.

Finally, it is worth emphasizing that the grand-canonical and canonical formalisms represent two physical extremes with respect to how surface charge responds to nuclear vibrations. The grand-canonical approach assumes instantaneous charge redistribution in response to atomic motion, whereas the canonical approach assumes the surface charge remains fixed throughout. In reality, the experimental situation is expected to lie between these two limits, depending on the time scale of electronic relaxation relative to vibrational dynamics. Importantly, our calculations reveal that the difference between these two limits becomes more significant as the surface coverage increases. At low coverage, the Stark tuning rates under grand-canonical and canonical ensembles are nearly indistinguishable, implying that the time-scale issue has a negligible impact and the experimental observable is robust. However, at higher coverage, the divergence between the two conditions grows, and the timing of charge redistribution becomes more critical in determining the Stark response. This underscores the fact that both the degree of interfacial molecular ordering and the dynamical nature of charge transfer must be considered when comparing theoretical models with experimental Stark tuning rates.

\section{Summary and conclusion}

In this work, we mainly worked on the grand-canonical ensemble on continuum model of implicit solvent and electrolyte effect. An automatic grand-canonical self-consistent DFT calculation loop is implemented with Quantum ESPRESSO code and Quantum ENVIRON code, which is available in the latest version of these two codes. We also proved, by using the Legendre transform, that the atomic forces in grand-canonical ensemble is identical to that of canonical ensemble, which is a normal DFT calculation do. And thus normal Hellmann-Feymann forces calculated in a normal DFT code can be used in a grand-canonical ensemble. On the other hand, by taking further derivatives of the Legendre-transformed energy expressions with respect to atomic displacements, we identify a correction term in the form of $\left.\left(\frac{\partial N_{e}^{*}\left(\alpha,E_{f}\right)}{\partial\alpha_{J\nu}}\right)_{E_{f}}\right|_{E_{f}^{*}}\cdot
\left.\left(\frac{\partial E_{f}^{*}\left(\alpha,N_{e}\right)}{\partial\alpha_{I\mu}}\right)\right|_{N_{e}^{*}}$ which quantifies the difference between vibrational frequencies computed under the grand-canonical and canonical ensembles. Based on this formal difference, and utilizing the established relationship between surface capacitance and chemical hardness, we derive an analytical expression linking the force constant difference between the two ensembles. This allows us to perform calculations in one ensemble and accurately estimate the vibrational frequencies in the other. Furthermore, by invoking the concept of chemical hardness at electrochemical interfaces, we demonstrate that the difference in force constants between the ensemble and grand-canonical descriptions converges for sufficiently thick metal slabs. In the meantime, this difference scales inversely with the surface area of the metal.

We performed finite difference calculations of vibrational frequencies and Stark tuning rates, using a carbon monoxide molecule adsorbed on a platinum slab surface as a representative example to illustrate the differences between canonical and grand-canonical treatments. After a thorough convergence test on both DFT and finite difference parameters, we observed that the vibrational frequency exhibits a significant difference when the vibrational mode is oriented perpendicular to the metal surface—where atomic displacement induces a large change in surface charge. In contrast, when the vibrational mode lies parallel to the surface, the frequency difference is much smaller, despite also involving notable charge redistribution. Moreover, our data support that the discrepancy between grand-canonical and canonical results diminishes when the occupation rate is small. This is because the sensitivity of surface charge to atomic displacement decreases as the surface area grows. Furthermore, based on our numerical results, we confirm that vibrational frequencies under the grand-canonical ensemble can be accurately predicted using only canonical DFT calculations combined with the analytically derived correction terms. The close agreement between the finite-difference results and our analytical predictions provides validation of the theoretical formula.

We extended our analysis to the calculation of Stark tuning rates, aiming to compare theoretical predictions with experimental measurements. By adjusting the dielectric constant in the implicit solvent model to values that are closer to the effective permittivity of interfacial water, as suggested by experimental and molecular dynamics studies, we found that the computed Stark tuning rates better match the experimental values. In addition to the dielectric constant, we also explored the role of the interfacial distance in the implicit solvent model. We found that tuning this distance can further influence the computed Stark tuning rate, providing an additional degree of freedom to calibrate the model against experimental observations.

While a single parameter set reproducing both the experimental surface capacitance and the Stark tuning rate has not yet been identified, this does not diminish the utility of the model. Rather, it indicates that some degree of flexibility or further calibration may be beneficial depending on the property of interest. Importantly, however, these uncertainties in solvent model parameterization do not affect the core conclusions of our work concerning the differences in vibrational frequencies under canonical and grand-canonical ensembles.

More broadly, our analysis highlights an important conceptual point: when computing response properties—such as forces, force constants, or polarizabilities—under grand-canonical ensemble, one must account for the implicit dependence of electronic degrees of freedom—particularly charge and Fermi level—on the variable being differentiated, such as atomic displacements. Neglecting such coupling can lead to inaccurate derivatives and misinterpretation of potential-dependent behaviors. Therefore, while total energies and forces may remain equivalent between canonical and grand-canonical formulations, second or higher-order derivatives often require explicit correction terms that reflect the ensemble's variable electron number.
\section{\label{sec:level1}Acknowledgement}
We acknowledge support by the NCCR MARVEL, a National Centre of Competence in Research, funded by the Swiss National Science Foundation (grant number 205602).
O.A.thanks the NSF CAREER award \#2306929. We also acknowledge the NSF CyberTraining award number 2321102 and, in particular, the 2023 Q-MS Rutgers Hackathon that have enabled the development of a significant part of the software presented in this work.

\bibliography{cite}
\end{document}